\documentstyle[sprocl]{article}

\def\be{\begin{equation}}
\def\ee{\end{equation}}
\def\bea{\begin{eqnarray}}
\def\eea{\end{eqnarray}}
\def\ifmath#1{\relax\ifmmode #1\else $#1$\fi}%
\def\rd{\ifmath{{\mathrm{d}}}}
\def\re{\ifmath{{\mathrm{e}}}}
\def\rg{\ifmath{{\mathrm{g}}}}
\def\rJ{\ifmath{{\mathrm{J}}}}
\def\rp{\ifmath{{\mathrm{p}}}}
\def\rP{\ifmath{{\mathrm{P}}}}
\def\rt{\ifmath{{\mathrm{t}}}}
\def\rT{\ifmath{{\mathrm{T}}}}

\def\rq{\ifmath{{\mathrm{q}}}}
\def\rem{\ifmath{{\mathrm{em}}}}
\def\eff{\ifmath{{\mathrm{eff}}}}

\def\diff{\ifmath{{\mathrm{diff}}}}

\def\resummed{\ifmath{{\mathrm{resummed}}}}
\def\BFKL{\ifmath{{\mathrm{BFKL}}}}

\begin{document}

%To Prof Nick Karayiannis -- do read this:-
%If needed the word of Chapter~1, you can type in at the 
%\title{}. The words will be in caps and lowercase. 
%For chapter title can be in all caps or in caps and lowercase.
%It is up to the author to type for the case sensitive but 
%all articles must be in the same style. 
%But mostly for Review Volume are without this Chapter~1.
%Thank you
%Jessie   13/4/2000

\title{INTRODUCTION TO LOW $x$ PHYSICS AND DIFFRACTION \footnote{Introductory 
talk presented at 
 the  30th International Symposium on Multiparticle Dynamics (ISMD 2000), 9-15 Oct 2000, Tihany, 
   Lake Balaton, Hungary.}}

\author{J. Kwieci\'nski}

\address{Department of Theoretical Physics, H. Niewodnicza\'nski Institute of Nuclear Physics, 
\\ Krak\'ow, Poland \\E-mail: jkwiecin@solaris.ifj.edu.pl} 

%\author{A. N. OTHER}

%\address{Department of Physics, Theoretical Physics, 1 Keble Road,\\
%Oxford OX1 3NP, England\\E-mail: other@tp.ox.uk}


\maketitle\abstracts{The basic concepts relevant for the theoretical
description of deep inelastic scattering within the QCD improved
parton model are introduced.  Recent developments in low $x$ DIS and
in deep inelastic diffraction are briefly summarised.  This includes
discussion of the BFKL dynamics including the subleading effects and
of the saturation model.  The dedicated measurements, which probe the
QCD pomeron are also discussed. }
%\documentstyle[12pt]{article}
%\begin{document}

%\section *{Introduction}.  
The  aim of this  talk is to discuss the following issues: 
\begin{enumerate}
\item Low $x$ physics in QCD. 
\begin{description}
\item {(a)} BFKL equation. 
\item{(b)}  Saturation model. 
\end{description}
\item Deep inelastic diffraction. 
%\item Summary and conclusions. 
\end{enumerate}

We shall discuss the low $x$ physics in QCD on the example of 
deep inelastic ep scattering, i.e. the process:  

\be
\re(p_\re) ~ + ~\rp (p ) \rightarrow \re(p_\re^{\prime}) + X\ .
\ee

The conventional kinematical variables for the description of this process 
are
\be
s=(p_\re + p)^2,~~~~ q=p_\re - p_\re^{\prime},~~~~ Q^2=-q^2,~~~~ 
W^2=(q+p )^2
\ee

\be
y={p q\over p_\re p },~~~~x={Q^2\over 2p q}\ .
\ee

The ep inelastic scattering is controlled by the virtual photon exchange 
mechanism and the total $\gamma^*\rp$ cross-section is closely related to the 
structure function $F_2$
$$\sigma_{\gamma^*\rp} = {4\pi^2 \alpha_{\rem}\over Q^2} F_2\ .$$
It should be observed that the small $x$ behaviour of 
$F_2$ is related to the  large $W^2$, i.e. Regge limit of 
$\sigma_{\gamma^*\rp}$. 

The ep DIS is conventionally described within the 
QCD improved parton model.  In this model, the structure function 
 $F_2$ is directly related to the quark and antiquark distributions 
 in the nucleon: 
 
\be
F_2(x,Q^2) = x \sum_f e_f^2 [q_f(x,Q^2) + \bar q_f(x,Q^2)] + O(\alpha_s)\ .
\ee

The parton (i.e. quark and gluon) distributions satisfy 
the DGLAP equations, which in LO have the following structure: 

$$
Q^2{\rd q_i  (x,Q^2)\over \rd Q^2} =  {\alpha_s(Q^2)\over 2\pi} 
( P^{(0)}_{\rq\rq} \otimes q_i   +  P^{(0)}_{\rq\rg} \otimes g)
$$

\be
Q^2{\rd g (x,Q^2)\over \rd Q^2} =
{\alpha_s(Q^2)\over 2\pi} 
[ P^{(0)}_{\rg\rq} \otimes \sum_i(q_i +\bar q_i)+P_{\rg\rg} \otimes g]\ .
\ee

Beyond LO (i.e. at NLO + ...) we have:  

$${\alpha_s(Q^2)\over 2\pi}P^{(0)}_ {ij} \rightarrow $$ 
\be
 {\alpha_s(Q^2)\over 2\pi}P^{(0)}_ {ij} + 
\left({\alpha_s(Q^2)\over 2\pi}\right)^2 P^{(1)}_ {ij}+....
\ee

At low $x$, the dominant role is played by the gluons. That follows from the 
singular behaviour of the splitting function $P_{\rg\rg}(z)$ for 
$z\rightarrow 0$,\\
$$P^{(0)}_{\rg\rg}(z) \sim {2 N_c\over z}\ .$$
The small $x$ behaviour of the parton distributions depends upon the structure 
of their small $x$ behaviour at the reference scale $Q_0^2$, i.e. 

$$xp_i(x,Q_0^2) \sim x^{-\lambda} \rightarrow xp_i(x,Q^2) \sim x^{-\lambda}$$
\be
\lambda > 0
\ee

\be
xp_i(x,Q_0^2) \sim const \rightarrow xp_i(x,Q^2) \sim 
\exp[2\sqrt{\xi(Q^2)\ln(1/x)}]\ ,
\ee
where
$$\xi(Q^2) = \int_{Q_0^2}^{Q^2} {\rd q^2\over q^2} 
{N_c\alpha_s{q^2}\over \pi}\ .$$

In both cases, 
$xg,~~~ x(q +\bar q),~~~ F_2... $ etc. are found to increase in the limit 
$x \rightarrow 0$.

The LO and NLO, the DGLAP formalism which sums the leading and 
next-to-leading powers of $\ln(Q^2/Q_0^2)$  is incomplete 
at low $x$.  In this region one has to resum (leading and subleading) powers 
of $\ln(1/x)$.  
 Small-$x$ resummation of leading (+subleading) powers of 
 {\bf $\ln(1/x)$} generates the QCD pomeron. \\
%(Balitzkij, Fadin, Kuraev, Lipatov (BFKL)).\\

 Diagramatically, the QCD Pomeron corresponds to  gluon ladder exchange.\\
The basic dynamical quantity in this case is the unintegrated gluon 
distribution $f(x,\hat k^2)$ where $\hat k^2$ denotes the square of 
the transverse momentum of the gluon.  
The unintegrated gluon distribution satisfies the 
Balitsky-Fadin-Kuraev-Lipatov (BFKL) equation, which, in the leading 
$\ln(1/x)$ approximation, has the following form:\cite{BFKL,GLR} 
\be
f(x,\hat k^2) = f^0(x,\hat k^2) + {3 \alpha_s\over \pi} K\otimes f \ ,
\ee

where
\be
K\otimes f = \int_x^1 {\rd z\over z}\int {\rd^2\hat q\over  \pi \hat q^2} 
[f(z,(\hat k + \hat q)^2) - \Theta(\hat k^2-\hat q^2)f(z,\hat k)]\ .
\ee

The conventional (integrated) gluon ditribution is given by: 
\be
xg(x,Q^2) = \int^{Q^2} {\rd\hat k^2\over \hat k^2} f(x,\hat k^2)\ .
\ee
 
The following properties of the BFKL dynamics should be mentioned:   

\begin{description}
\item{1.} Diffusion of transverse momentum along the chain 
which should reflect itself in the hadronic final state. 
\item{2.} Characteristic rise with decreasing $x$.\\ 

In LO $f \sim x^{-\lambda}, \lambda = 4\ln(2) 3\alpha_s/\pi$\\
\item{3.} Large subleading effects. Their major part is understood and is 
under control.
\item{4.} The BFKL equation embodies (part of the) LO DGLAP evolution. \\
\end{description}
              
One of the most important recent theoretical developments in the low $x$ 
physics has been the completion of the calculation of the NLO $\ln(1/x)$ 
effects.\cite{BFKLNL1,RESUMCOL}
%(V. Fadin, L. Lipatov, M.Kotski, Fiore, M. Ciafaloni, D. Colferai, G. Salam.......)
%\vspace{0.5cm}
Those effects were found to be very important and in particular they 
were found to reduce significantly the QCD pomeron intercept, which in 
the NLO approximation is given by                       
\be
\lambda = 4\ln(2)\bar \alpha_s(1-6.3\bar \alpha_s)\ ,
\ee

where

$$\bar \alpha_s = {3\alpha_s\over \pi}\ . $$

It is, therefore, obvious that resummation of subleading $\ln(1/x)$ 
beyond NLO is needed. It has been found, however, that the dominant part 
of the subleading corrections 
is generated by the  phase-space limitations. They can be taken into account  
 exactly, i.e. beyond NLO.\cite{RESUMCOL,KMSG,KMS}\\ 
%
%(M. Ciafaloni, D. Colferai, G. Salam....., 
%B. Anderson, G. Gustaffson, J. Samuelsson, D. Karraziha, ...
%JK, A.D. Martin, A. Stasto, P. Sutton)
%

The observable quantities, as the structure function $F_2$ are obtained 
from the unintegrated gluon distributions through 
the $k_\rt$ factorisation:\cite{KTFAC} 
\be
F_2 \sim F^{\gamma g} \otimes f \ ,
\ee 
where '$\otimes$' denotes in this case convolution in transverse and 
longitidinal momenta. At leading twist, the $k_\rt$ factorisation theorem 
can be recast into conventional collinear factorisation form: 
\be
Q^2 {\partial F_2 \over \partial Q^2} 
\sim {\alpha_s\over 2 \pi}P_{\rq\rg}^{\resummed}
(\alpha_s)\otimes xg^{\BFKL}\ ,
\ee 
where the leading (and possibly also subleading) $\ln(1/x)$ effects are 
included to all orders in $xg^{\BFKL}$ and in the splitting function 
$P_{\rq\rg}^{\resummed}$.  
They include in particular (part of) conventional DGLAP NNLO effects.   
Small-$x$ resummation in $P_{\rq\rg}^{\resummed}$ has important 
implications for the extraction of $xg$ from the scaling violations.\cite{PQG} 
The recent NNLO analysis of the DGLAP equations which embodies 
those effects shows that the gluon distributions in NNLO 
approximation are significantly smaller at small $x$ 
than those obtained within the NLO framework.\cite{NNLO}\\
%(F. Catani, R.K. Ellis, Z. Kunszt, F. Hautmann, B.R. Webber, E. Levin,...
%R. Thorne, ...JK, A.D. Martin, A. Stasto,......)  

It is possible to obtain a very economical description 
of the $F_2$ HERA data within the BFKL - $k_\rt$ factorisation 
framework.\cite{KMS}  
One can, however, get an equally good description of the data staying 
within the 
conventional NLO DGLAP formalism.  The measurement of the structure function 
alone is, therefore, not a sensitive discriminator of the underlying dynamics. 
In order to probe the details of the QCD pomeron, it is particularily useful 
to study the high energy processes characterised by two comparable scales 
$Q_1^2$ and $Q_2^2$  at the 'ends' of the gluon ladder which corresponds 
to the QCD pomeron. In this kinematical configuration, the conventional 
LO DGLAP evolution from the scale $Q_1^2$ to $Q_2^2$ is suppressed and the 
corresponding cross-sections are sensitive to the diffusion of the 
transverse momenta along the chain, which is a chacteristic feature 
of the BFKL dynamics.  The following dedicated  measurements are 
particularily useful for this purpose:\\ 

$\bullet~~$ Two-jet production in high 
energy hadronic collisions with $k_{\rT1}^2 \sim k_{\rT2}^2$.{\cite {MNAV}\\  

$\bullet~~$ Forward jet ($k_{j\rT}^2 \sim Q^2$)  (or forward $\pi^0$) 
production in ep DIS.\cite{FJET}\\
%(A.H. Mueller, H. Navelet,.... A. De Roeck, J. Bartels, S. Munier, JK, 
%A.D. Martin, P. Sutton, J. Outhwaite.....)
                        
$\bullet~~$ Doubly tagged e$^+$e$^-$ events which are related to the 
$\gamma^{*}(Q_1^2) \gamma^{*}(Q_2^2)$ total 
cross-section.\cite{GSTGST,L3GG,OPAL} \\
%(J. Bartels, A. De Roeck, C. Ewerz, H. Lotter,... S. Brodsky, F. Hautmann, 
%D. Soper,... A. Bia\l{}as, W. Czy\.z, W. Florkowski,.... M. Boonekamp, C. Royon, 
%A. De Roeck,  
%S. Wallon,..... A. Donnachie, 
%H. Dosch, M. Rueter,.... N. Nikolaev, J. Speth, V. Zoller,..... E. Gotsman, 
%E. Levin, U. Maor, E. Naftali,.....V. Kim, L. Lipatov,....JK, L. Motyka...) 

Indefinite increase of parton distributions $xp(x,Q^2)$ with decreasing 
$x$ cannot hold forever. The QCD improved parton 
model based upon  linear evolution equations has to break down when  
$${xp(x,Q^2)\over Q^2} \sim \pi R^2\ ,$$ 
where $R$ denotes the (transverse) radius describing the size of the region 
within which the partons are concentrated.  
In the small $x$ region, the linear evolution equations have to be modified 
by the non-linear screening corrections which eventually lead to parton 
saturation.\cite{GLR,SHAD} 
%(L. Gribov, L. Levin, M. Ryskin...., A. Mueller, J. Qiu...., J. Bartels, G. Schuler, 
%J. Bl\"umlein, ....J. Bartels,  
%C. Ewerz,.....M. Armesto, M. Braun,.....Yu. Kovchegov,....M. Gay-Ducati, V. Goncales...., 
%L. MacLerran, R. Venugopalan,....J. Jalilian-Marian, 
%A. Kovner, A. Leonidov, H. Weigert,...)
%\vspace{1cm}                       
%$$ xp(x,Q^2) \rightarrow xp^{eff} \sim \pi R^2 Q^2$$ 
A semi-phenomenological approach to saturation has recently been developed 
within the colour dipole model by K. Golec-Biernat and  
M. W\"usthoff.\cite{KGBMW1,KGBMW2}
This formulation, that has proved to be phenomenologically very successful, 
utilises the picture in which the high energy 
$\gamma^*\rp$ total cross-section is driven by the interaction of the 
$\rq \bar\rq$ colour dipole into 
which the virtual photon fluctuates,\cite{SHAD} i.e.                  
\begin{equation}
\sigma_{\gamma^*\rp}(Q^2,x) \sim \int \rd z \rd r^2
|\Psi(r,Q,z)|^2\sigma_{\rq \bar\rq}(r,x)\ .  
\label{dipole}
\end{equation}
 In equ.~(\ref{dipole}),  
 $\Psi(r,Q,z)$ denotes   the wave function of the virtual photon,
 $ \sigma_{\rq \bar\rq}(r,x)$ is  
the total cross-section describing the interaction of the $\rq \bar\rq$ dipole 
with the proton target, $r$ is the transverse size of the dipole and $z$ is 
the momentum fraction of the virtual photon carried by a quark (antiquark). 
In the leading $\ln(1/x)$ approximation, the dipole picture corresponds to the 
$k_\rt$ factorisation formula (partially) transformed into the coordinate 
representation.\cite{BNP}
In the formulation of the model discussed in ref. \cite{KGBMW1},
we have               
$$\sigma_{\rq \bar \rq}(r,x)=\sigma_0[1-\exp(-r^2/R_0^2(x))]\ ,$$
where the saturation radius $R_0(x)$ is a decreasing function of 
the parameter $x$ and is parametrised in the following form:
\begin{equation} 
R_0^2(x) \sim x^{\lambda} \ .
\label{rzero}
\end{equation}
The parameter $\sigma_0$ denotes the magnitude of the dipole cross-section 
in the large $r$ limit.  The behaviour of the $\gamma^*\rp$ cross-section 
(\ref{dipole}) in the region of large (small) values of $Q^2$ is linked 
with the properties of the dipole cross-section $\sigma_{\rq \bar\rq}(r,x)$ 
for small (large) values of the dipole size $r$. To be precise, we 
have:\cite{KGBMW1}
 $$r^2\ll R_0^2(x) \leftrightarrow Q^2 \gg 1/R_0^2(x)\ ,$$
where  
\begin{equation}
\sigma_{\rq \bar\rq}(r,x) \sim r^2/R_0^2(x)\ , \label{smr}
\end{equation}
that gives 
 \begin{equation}
\sigma_{\gamma^*\rp}(Q^2,x)  \sim {1\over Q^2R_0^2(x)}
\label{lq2}
\end{equation} 
%Saturation of $\sigma_{\rq \bar \rq}(r,x)$ and $\sigma_{\gamma^*p}(Q^2,x)$:\\ 
and 
$$r^2>R_0^2(x) \leftrightarrow Q^2 <1/R_0^2(x)\ ,$$
where   
\begin{equation}
\sigma_{\rq \bar\rq}(r,x) \sim \sigma_0\ ,  
\label{lr}
\ee
that gives 
\begin{equation}
\sigma_{\gamma^*\rp}(Q^2,x)  \sim \ln[Q^2R_0^2(x)]\ .
\label{smq2}
\end{equation}
The latter behaviour corresponds to the saturation of the cross-section. 
The remarkable property of the saturation model  is geometric scaling of 
$\sigma_{\gamma^*\rp}(Q^2,x)$,\cite{GS} which means that at low values of $x$ 
this cross-section becomes the function of only one dimensionless variable, 
i.e. 
% (A. Sta\'sto, K. Golec-Biernat, JK)
\begin{equation}                         
\sigma_{\gamma^*\rp}(Q^2,x) \rightarrow \Phi(\tau)\ ,
\label{gs}
\end{equation}
where 
\begin{equation} 
\tau = Q^2R_0^2(x)\ .
\label{tau}
\end{equation}
The geometric scaling (\ref{gs}) is very well supported by experimental 
data.\cite{GS} \\
              
Deep inelastic diffraction in ep inelastic scattering is a process: 
\begin{equation}
\re(p_\re)+\rp(p) \rightarrow \re^{\prime}(p_\re^{\prime}) + X 
+\rp^{\prime}(p^{\prime})\ ,
\label{disdif}
\end{equation}
where there is a large rapidity gap between the recoil proton 
(or excited proton) and the hadronic system $X$.\cite{DIFF,ADMMW}
To be precise,
process (\ref{disdif}) reflects the diffractive disssociation 
of the virtual photon.  Diffractive dissociation is described by the following 
kinematical variables: 
\begin{equation}
\beta={Q^2\over 2 (p-p^{\prime})q}
%\nonlabel
\end{equation}
\begin{equation}
x_\rP={x\over \beta}
%\nonlabel
\end{equation}
 \begin{equation}
t= (p-p^{\prime})^2. 
\label{difv}
\end{equation}
Assuming that diffraction dissociation is dominated by the pomeron 
exchange and that the pomeron is described by 
a Regge pole, one gets the following factorizable expression for the 
diffractive structure function:\cite{FACREG}
\begin{equation}
{\partial F_2^{\diff}\over \partial x_\rP \partial t}= 
f(x_\rP,t)F_2^\rP(\beta,Q^2,t)
\label{difsf}
\end{equation}
where the "flux factor" $f(x_\rP,t)$ is given by the following formula
:
\begin{equation}
f(x_\rP,t)=N{B^2(t)\over 16\pi} x_\rP^{1-2\alpha_\rP(t)}\ ,
\label{flux}
\end{equation}
with $B(t)$ describing the pomeron coupling to a proton and $N$ being the 
normalisation factor.  The function $F_2^\rP(\beta,Q^2,t)$
is the pomeron structure function, which, in the (QCD improved) parton model,
is related in a standard way to the quark and antiquark distribution 
functions in a pomeron:
\begin{equation}
F_2^\rP(\beta,Q^2,t)=\beta \sum e_i^2[q_i^\rP(\beta,Q^2,t)+ \bar 
q_i^\rP(\beta,Q^2,t)]\ ,
\label{f2pom}
\end{equation} 
with $q_i^\rP(\beta,Q^2,t)=\bar q_i^\rP(\beta,Q^2,t)$.  The variable 
$\beta$, which is the Bjorken scaling variable appropriate for 
deep inelastic lepton-pomeron "scattering",  has the meaning of the 
momentum fraction of the pomeron carried by the 
probed quark (antiquark). The quark and gluon  distributions in a pomeron 
are assummed to obey the standard Altarelli-Parisi evolution equations: 
The deep inelastic diffraction may therefore probe the quark-gluon content 
of the Pomeron. \\  

The deep inelastic diffraction is sensitive to the interplay between the 
soft and hard pomerons.\cite{BARTKOW} It turns out that the effective 
pomeron intercept extracted from 
the diffractive data is higher than that of the soft pomeron , i.e. 
(effective) $\alpha_{\rP}^{\eff}(0)\sim 1.2$. This implies an important 
contribution of the hard pomeron exchange. One also finds important higher 
twist contribution to the diffractive structure functions.\\

Important diffractive processes which can probe 
the hard pomeron are the following ones:\cite{ADMMW}\\ 

 $\bullet~\gamma^{*}+ \rp \rightarrow V+\rp$\\

$\bullet~(\gamma, \gamma^{*})+ \rp \rightarrow \rJ/\Psi+\rp$ \\ 

 $\bullet~$ diffractive jet production\ .\\

One of the  still open problems of the theory of hard diffraction 
processes is the violation of the QCD (collinear) factorisation 
in hadronic collisions.\cite{FACTB}
Finally, let us point out that within the saturation model the total 
diffractive cross-section is given by:  
\be
\sigma^{\diff}_{\gamma^*\rp}(Q^2,x) \sim \int \rd z \rd r^2
|\Psi(r,Q,z)|^2 \sigma^2_{\rq \bar\rq}(r,x) \ .
\ee
The fact that $\sigma^{\diff}_{\gamma^*\rp}(Q^2,x)$ is given in terms of 
 $\sigma^2_{\rq\bar\rq}(r,x)$ implies that it is sensitive to the contribution 
from the large $r$ region. \\

To sum up, we have introduced in this talk the basic concepts of low
$x$ physics and of hard diffraction.  We have also summarised some of
the most recent developments including NLO BFKL, the saturation model,
etc. Theoretical QCD predictions for low $x$ phenomena have been
intensively studied both at HERA and Tevatron \cite{EXPER} as well as
at LEP.\cite{L3GG,OPAL}  Most of those predictions are also extremely
relevant for the measurements at future colliders, which will open up
hitherto unexplored regime(s).
              
\section*{Acknowledgments}     
I thank Albert De Roeck and Dino Goulianos for their kind invitation to Tihany 
and I congratulate Thomas Cs\"org\H{o} and Wolfram Kittel for organising 
an excellent meeting.  
This research has been partially  supported by the EU Framework 
TMR programme, contract FMRX-CT98-0194.

\end{document}